\documentclass[manuscript]{aastex}
\shortauthors{emulateapj}
\usepackage{natbib}
\usepackage{amsmath,amssymb,verbatim}
\usepackage{lipsum}
\usepackage{color}

\begin{document}
\title{Black Hole Mass Estimates and Rapid Growth of Supermassive Black Holes in Luminous $z \sim$ 3.5 Quasars} 

\author{Wenwen Zuo\altaffilmark{1, 3}, Xue-Bing Wu\altaffilmark{1, 2}, Xiaohui Fan\altaffilmark{2, 3}, 
Richard Green\altaffilmark{3}, Ran Wang\altaffilmark{2}, \& Fuyan Bian\altaffilmark{4}} 

\altaffiltext{1}{Department of Astronomy, School of Physics, Peking University, Beijing 100871, China} 
\altaffiltext{2}{Kavli Institute for Astronomy and Astrophysics, Peking University, Beijing 100871, China}
\altaffiltext{3}{Steward Observatory, The University of Arizona, Tucson, 85721, USA}
\altaffiltext{4}{Research School of Astronomy \& Astrophysics, Mount Stromlo Observatory, Cotter Road, 
Weston ACT 2611, Australia}

\begin{abstract}
We present new near-infrared (IR) observations of the H$\beta\ \lambda4861$ and MgII $\lambda2798$ lines for 32 luminous quasars with $3.2<z<3.9$ using the Palomar Hale 200 inch telescope and the Large Binocular Telescope. We find that the MgII Full Width at Half Maximum (FWHM) is well correlated with the H$\beta$ FWHM, confirming itself as a good substitute for the H$\beta$ FWHM in the black hole mass estimates. The continuum luminosity at 5100 \AA\ well correlates with the continuum luminosity at 3000 \AA\ and the broad emission line luminosities (H$\beta$ and MgII). With simultaneous near-IR spectroscopy of the H$\beta$ and MgII lines to exclude the influences of flux variability, we are able to evaluate the reliability of estimating black hole masses based on the MgII line for high redshift quasars. With the reliable H$\beta$ line based black hole mass and Eddington ratio estimates, we find that the $z\sim3.5$ quasars in our sample have black hole masses $1.90\times10^{9} M_{\odot} \lesssim M_{\rm BH} \lesssim 1.37\times10^{10} M_{\odot}$, with a median of $\sim 5.14\times10^{9} M_{\odot}$ and are accreting at Eddington ratios between 0.30 and 3.05, with a median of $\sim1.12$. Assuming a duty cycle of 1 and a seed black hole mass of $10^{4} M_{\odot}$, we show that the $z\sim3.5$ quasars in this sample can grow to their estimated black hole masses within the age of the Universe at their redshifts. 
\end{abstract}


\section{INTRODUCTION\label{introduction}}
Active Galactic Nuclei (AGNs) are generally accepted to be powered by the gravitational potential 
energy extracted from matter falling towards a supermassive black hole (SMBH). The scaling relations 
between central BH masses and various properties of their host galaxy spheroidal components,  such as bulge mass, luminosity and stellar velocity dispersion, strongly suggest that BH growth is coupled with galaxy formation and evolution \citep{Gebhardt00, Ferrarese00, Onken04, Nelson04, Kormendy13}. The observed luminosity function and BH mass function at different redshifts reveal a downsizing phenomenon, whereby more massive black holes underwent their active accretion in earlier cosmic time; the comoving number density of massive active BHs peaks at earlier time than the less massive active BHs \citep{Ueda03, Richards06, Vestergaard08, Vestergaard09, Kelly10, Shen_Kelly10, Shen_Kelly12, Kelly12, Kelly13}. Combination of the BH mass function with the luminosity function can better constrain the growth of SMBHs and their connections with the galaxy evolution \citep{Vestergaard08, Kelly12}. Reliable BH mass estimates are of fundamental importance in determining the BH mass functions and understanding other physical processes related to BHs.  

In the local universe, BH mass estimates of Seyfert 1 galaxies and quasars have been done successfully with the reverberation mapping (RM) technique, mainly involving the H$\beta$ line \citep[e.g.][]{Kaspi00, Peterson04, Bentz10, Barth11, Grier12, Grier13}. Based on these results, tight empirical correlations between the broad line region size and the continuum or emission line luminosity in optical wavelength have been established \citep{Kaspi00, Wu04, Peterson04, Kaspi05, Bentz06, Bentz09, Bentz13}. Assuming that the broad line region gas is virialized \citep{Peterson99, Peterson00, Peterson04}, these correlations have been widely adopted to estimate the BH masses of low redshift AGNs with single-epoch optical spectra based on the H$\beta$ line \citep{Greene05, Vestergaard06, McGill08, Vestergaard09}. In the shorter ultraviolet (UV) wavelength range, such estimates have also been extended to the cases of  the MgII $\lambda2798$ or CIV $\lambda1549$ emission line based BH masses for small samples of AGNs \citep{Vestergaard02, McLure04, Warner04, Vestergaard06, Kong06, Kollmeier06}. These scaling relations are important for estimating  the BH masses of high redshift AGNs because the H$\beta$ line moves out of the optical window and we have to rely on the MgII and CIV emission lines for AGNs at $z>0.7$. The MgII and CIV lines have been adopted to estimate BH masses for a large sample of SDSS quasars and to study the BH growth history \citep{Vestergaard04, Shen08}. However, the reliability of the BH mass estimates based on the MgII or the CIV emission lines is still controversial, compared to the H$\beta$ line based estimates \citep{Baskin05, Netzer07, Trakhtenbrot12}. Some studies have confirmed the consistency \citep{Vestergaard06, Shen08, Wang09, Shen12}, but some proposed additional calibrations to reduce the deviations from the H$\beta$ line based BH mass estimates \citep{Denney12, Park12, Denney13, Park13}.

At high redshift, the H$\beta$ line shifts to the near-IR window. Thus, with the near-IR spectroscopy, we can still directly use the well established empirical relation involving the H$\beta$ line to obtain the BH masses for high redshift quasars.
\cite{Shemmer04}  and \cite{Netzer07} obtained $H$ and $K$ band spectroscopy of the H$\beta$ line and derived BH masses of $10^{8.8}$-$10^{10.7} M_\odot$ for 44 quasars with redshift between 2.2 and 3.4. \cite{Dietrich09} obtained $J$ and $H$ band spectroscopy for 10 quasars with redshift between 1.0 and 2.2 and derived H$\beta$ line based BH masses of $10^{9.3}$-$10^{10} M_\odot$. \cite{Greene10} used the Triplespec of the 3.5 m telescope at Apache Point Observatory to obtain near-IR spectroscopy of 16 lensed quasars with redshift between 1.0 and 3.6 (only one at $z>3$) and derived H$\beta$ and H$\alpha$ line based BH masses of $10^{8.8}$-$10^{9.9} M_\odot$. \cite{Trakhtenbrot11} presented $H$ band spectroscopy of 40 AGNs with redshift around 4.8 and obtained MgII line based BH masses of $10^{8}$-$10^{9.8} M_\odot$. \cite{Shen12} investigated the reliability of  UV virial mass estimates for a sample of 60 quasars  at $1.5<z<2.2$ and obtained H$\beta$ line based BH masses of $10^{8.5}$-$10^{10.5} M_\odot$. In addition, near-IR spectra have also been taken for some $z \sim$ 6 quasars, and  BH masses of $10^{8.5}$-$10^{9.7} M_\odot$ were obtained using the MgII and CIV lines \citep{Willott03, Jiang06, Kurk07, De11}. 

From these studies we can see that there still exist some redshift gaps ($z \sim 3.5$) without reliable H$\beta$ line based BH mass estimates. More importantly, based on multi-epoch spectra of 615 high redshift quasars selected from the SDSS, the uncertainties of BH mass estimates due to inherent variability are estimated to be $\sim$ 30\% \citep{Wilhite07}. All of the previous studies investigated only a single emission line (either H$\beta$ or MgII) in the near-IR bands; simultaneous high-quality observations of both H$\beta$ and MgII lines for $z\sim3.5$ quasars have rarely been done. However, this is essential in obtaining a reliable calibration between the H$\beta$ and MgII line based BH masses at high redshift; in this case, the only factor influencing the calibration is the intrinsic difference between these two simultaneously observed emission lines. Therefore, with this goal in mind, we have observed a sample of quasars with $3.2<z<3.9$ using the Palomar Hale 200 inch telescope and the Large Binocular Telescope. Our spectroscopy in $J$, $H$ and $K$ bands from 0.8-2.2 $\mu$m covers the MgII, H$\beta$ and FeII lines simultaneously, allowing us to critically examine the reliability of the BH mass estimates based on the MgII line.

In this paper, we present a brief overview of our sample, observations and data reduction in \textsection~\ref{Sample} and describe the method of spectral fitting in \textsection~\ref{Fitting}. We investigate correlations between spectral properties in \textsection~\ref{Result} and present modified calibration results based on the rest-frame UV emission lines in \textsection~\ref{Calibration}. The results are discussed in \textsection~\ref{Discussion}. The main results are summarized in \textsection~\ref{Conclusion}. We adopt $\Omega_{\Lambda} = 0.7$, $\Omega_{0} = 0.3$ and $H_{0} = \rm 70\ km\ s^{-1}\ Mpc^{-1}$ throughout the paper.

\section{SAMPLE, OBSERVATIONS AND DATA REDUCTION\label{Sample}}
We select our targets from the SDSS DR7 quasar catalog \citep{Shen11, Schneider10} mainly by constraining the redshift and magnitude ranges. The redshift range ($3.2 < z < 3.8$) ensures that both the H$\beta$ and MgII lines are well covered in our near-IR spectroscopy. Certain redshift ranges where the H$\beta$ or MgII lines fall in telluric absorption bands in the near-IR windows are avoided. We further restrict the Two Micron All Sky Survey (2MASS) \citep{Skrutskie06} $J$ and $K$ magnitude to brighter than 17.0 and 16.0 mag, respectively. 

As part of the Telescope Access Program in China{\footnote{http://tap.bao.ac.cn}}, 30 targets were observed
during 2010-2012 with the TripleSpec instrument mounted on the Palomar Hale 200 inch telescope.
TripleSpec is a cross-dispersed near-IR spectrograph providing continuous spectral coverage of 
0.95-2.46 $\mu$m simultaneously at a resolution of $\sim$ 2700 \citep{Terry08}. A slit width of 1$''$ was used
during observations, and the targets were nodded along the slit to obtain good background subtraction. 
Another 2 quasars were observed with the LUCI 1 near-IR instrument \citep{Hill12} mounted on the 
Large Binocular Telescope (LBT).
LUCI 1 is a near-IR imager/spectrograph for the LBT with wavelength coverage of
0.85-2.4 $\mu$m ($zJHK$ bands) in imaging, long-slit and multi-object spectroscopy modes. The spectra were
taken with the N1.8 camera and the 210 l/mm $J$, 210 l/mm $K$ gratings, yielding a spectral resolution of 1.5 \AA.
The total integration time in each band for each target is typically 2400-3600 s. A0V type stars were 
observed either before or after our targets at similar airmass, to remove telluric features in the following data 
reductions. Our sample and the follow-up observations are summarized in Table~\ref{sample}, where the 
systemic redshifts measured from their [OIII] $\lambda 5007$ lines are also listed. 
For all these targets, there are good SDSS spectra covering CIV $\lambda 1549$ with mean signal to noise ratio (S/N) per spectral resolution element (3 pixels) greater than 10.

Reduction of the raw spectroscopic data from TripleSpec is carried out using the modified IDL-based Spextool3 package \citep{Cushing04}. After the flat field correction using the lamp spectra, observations in different nodded positions are pair-subtracted to remove most of the background. For each target, all the individual exposures are wavelength calibrated and extracted to be one-dimensional spectra, which are then combined to get the final averaged spectrum and telluric corrected. The sky background subtraction, flat correction and wavelength calibration of the raw spectra from the LBT have been done with the modified IDL longslit reduction package for NIRSPEC \citep{Becker09, Bian10}.
After this step, the one-dimensional spectra are extracted using the IRAF and then telluric corrected in the same way with that for the TripleSpec data. Special attention is paid to telluric correction, where each extracted quasar spectrum is divided by a telluric spectrum. The telluric spectrum is obtained from dividing the observed A0V stellar spectrum by a scaled and reddened model Vega spectrum according to the observed magnitudes of the A0V star. Before this step, the model Vega spectrum is convolved with a function that broadens the lines to the observed widths and smooths it to the observed resolution \citep{Cushing04}. The best obtained telluric spectrum is then visually examined, with removing those broad emission or absorption lines by manually scaling their equivalent widths. The absolute flux calibration is done by comparing the 2MASS $JHK$ band magnitudes with the synthetic $JHK$ band magnitudes. These synthetic magnitudes are obtained by convolving the quasar spectrum with the 2MASS response function provided by \cite{Cohen03}. The final spectra are also corrected for the Galactic extinction \citep{Cardelli89, Schlegel98} and the wavelength is redshift corrected to the rest frame.

\section{SPECTRAL MEASUREMENTS\label{Fitting}}
For each emission line (H$\beta$ and MgII) in the near-IR spectroscopy, we fit a pseudo-continuum to the wavelength range around the line. The pseudo-continuum consists of a power-law continuum and Fe II template without including the contribution from Balmer continuum, because the Balmer continuum is difficult to be constrained from the noisy spectra in the gap between the $J$ and $H$ bands. The exclusion or inclusion of the Balmer continuum does not significantly influence the decomposition of broad lines \citep{Shen12}. However, they also stated that it may affect the continuum luminosity measurement by $\sim$ 0.12 dex. Note that this factor is omitted in our estimate of  luminosity measurement uncertainties. After subtracting the pseudo-continuum, the emission lines are fitted with multiple Gaussians. The emission line fitting process is briefly described as follows (See \cite{Shen11} for more details).

We fit the wavelength range of 4700-5100 \AA\  with 2 Gaussians for the broad H$\beta$ component, 1 Gaussian for the narrow H$\beta$ component and 2 Gaussians for the [OIII] $\lambda4959$ and [OIII] $\lambda5007$ narrow lines. The line center offset from its theoretical value (namely the line shift) and the Full Width at Half Maximum (FWHM) of the H$\beta$ narrow component are set to be the same as those of [OIII] $\lambda5007$. The intensity ratio of the [OIII] doublets is fixed to the theoretical value of 3.0 $(f_{5007}/f_{4959} = 3)$. If needed, we introduce two additional Gaussians for the extended wings of the [OIII] $\lambda4959$ and [OIII] $\lambda5007$  lines.

The MgII emission line is not treated as a doublet, since the line splitting is not apparent enough to affect the broad line width measurements. We fit 2 Gaussians for the broad component and 1 Gaussian for the narrow component in the rest-frame wavelength range of 2700-2900 \AA. Though the local continuum below the MgII and H$\beta$ lines are modeled independently, the narrow component of the MgII line is tied to [OIII] $\lambda5007$ with the same FWHM and line shift. Similar to \cite{Shen11}, an upper limit of 1200 km/s is set to the FWHM of narrow component. The measured narrow component FWHM varies from 500 km/s to 1200 km/s.

For the CIV emission line in the SDSS optical spectra, we fit the wavelength range 1500-1600 \AA\ with 2 Gaussians for the broad component and 1 Gaussian for the narrow component. The narrow component is freely fitted with an upper limit of FWHM of 1200 km/s. Note that this method could potentially add scatters to the measured CIV FWHM.

An example of the continuum and emission line fitting is shown in Fig. 1. Similar to \cite{Shen12}, a Monte-Carlo approach is applied to estimate the uncertainties of the fitting parameters. For each object, 50 random mock spectra are created by introducing random Gaussian noises to the original spectrum; at each pixel in a given mock spectrum, the noise term is randomly drawn from a normal distribution with the observed flux density error as the standard deviation. We then fit the mock spectra with the same fitting strategy. The 1$\sigma$ dispersions of these measurements relative to the originally measured values are thus considered to be the corresponding uncertainties. As stated in \cite{Shen11}, the flux density uncertainty of the mock spectrum is increased compared to the original one, since they are added twice. Nevertheless, these mock spectra can still be considered to represent the wavelength-dependent noise properties. 

The fitting results, such as emission line widths and continuum luminosities are listed in Table~\ref{fitting-results}. For quasars labeled as `good' (see next section for details), the spectra together with their best-fit models of the MgII and H$\beta$ lines are shown in Figs. 2 and 3, respectively.

\section{CORRELATIONS OF SPECTRAL PROPERTIES \label{Result}}
After excluding 6 Broad Absorption Line Quasars (BALs) from the 32 quasars, we calculate the median S/N per spectral resolution element (3 pixels) for each quasar over the whole spectroscopy. It is difficult to identify line features from the spectrum with S/N less than 10. Based on the criteria, J232735.67-091625.6 is labeled as `poor' with S/N $\sim 8.1$. Another target J012403.77+004432.6 at redshift of 3.834 is also labeled as `poor' (despite its high median S/N as 70), because its MgII and H$\beta$ emission lines are located in the gap between $J$ and $H$ and the gap between $H$ and $K$ bands, respectively; the accurate determination of the line profiles is prevented under this circumstance. For the 3 quasars without good telluric corrections due to the bad weather during the observations, we label their quality as `median'.
Therefore, based on the remaining 21 quasars labeled as `good', we proceed to investigate the correlations between widths of different lines and between continuum or emission line luminosities at different wavelengths, as well as those between their virial products. 

To describe the rank correlation between different parameters, such as the FWHMs of different emission lines, we estimate the Spearman rank correlation coefficient $r1$ and the probability $p$ of its deviation from the null hypothesis that there is no relationship between the analyzed parameters. Correlations with $r1$ greater than 0.5 are referred as strong. Those with $0.3 < r1 \le 0.5$ are considered as intermediate. Those with $0.1 < r1 \le 0.3$ are referred as weak and those with $r1 < 0.1$ are considered as none or very weak. Correlations with the probability $p$ less than 0.05 (confidence level larger than 95\%) are considered as significant. As suggested by \cite{Wang09}, consistent results can be obtained from most of the regression methods with or without errors, including the ordinary least-squares (OLS) method, FITexy \citep{Press92}, the bivariate correlated errors and intrinsic scatter (BCES) regression method \citep{Akritas96}. To perform a more comprehensive regression analysis on the spectral properties, we apply a commonly used OLS method (without errors) and the BCES method (with errors) to our sample. Though there is no prior knowledge about which of the two variables is independent, the BCES method, treating the two variables symmetrically, can yield reasonable results \citep{Akritas96, Wang09}. Accompanied by the regression analysis, the Pearson product-moment correlation coefficient $r2$ is also calculated to illustrate the strength of the proposed regression relationship.
 
These correlation coefficients are calculated and the regression fittings are performed only for the targets labeled as `good', although the quasars labeled as `median' are also shown in Fig.~\ref{fwhm_total} and Fig.~\ref{l_total}. Especially for the correlations between properties of two different lines, quasars without full spectral coverage of both lines are excluded. Among the 21 quasars labeled as `good', 20 of them have full coverage of the H$\beta$ line, 20 of them have full coverage of the MgII line and all the 21 quasars have full coverage of the CIV line. 19 of them have full coverage of both the H$\beta$ line and MgII line.

\subsection{Line Width Correlations\label{Linewidth}}
Consistent with earlier studies \citep{Greene05, Wang09, Shen12}, we also find strong correlations between the FWHM of the broad H$\beta$ component (FWHM$_{\rm H\beta}$) and that of the broad MgII component (FWHM$_{\rm MgII}$), with $r1 \sim$ 0.72 at a confidence level over 99\%. As shown in Fig.~\ref{fwhm_total}, the slope from fitting their logarithmic values with the OLS method is 1.09$\pm$0.23. This confirms the FWHM of the MgII line as a good substitute for FWHM$_{\rm H\beta}$ in the BH mass estimates of our $z\sim3.5$ quasars \citep{Schneider10, Shen11}.
Combined with the SDSS spectroscopy of our sample, a moderate correlation is found between FWHM$_{\rm H\beta}$ and FWHM$_{\rm CIV}$ ($r1 \sim 0.29, p \sim 0.21$), suggesting that the FWHM of the CIV line can still be a proxy to estimate BH masses instead of FWHM$_{\rm H\beta}$. This weaker relationship between FWHM$_{\rm H\beta}$ and FWHM$_{\rm CIV}$ may be due to several factors, such as S/N in SDSS spectra, skewness of line measurements induced by absorption in the CIV profile, possible line profile variations between the SDSS and our near-IR observational epochs or the combined influences of these factors. Results from other correlation analyses are also shown in Table~\ref{fwhm}.  

\subsection{Luminosity Correlations\label{Luminosity}}
In Fig.~\ref{l_total}, we compare different luminosities with the continuum luminosity at 5100 \AA\ ($L_{5100}$).  The results are listed in Table~\ref{fwhm}. Because our objects are luminous ($10^{47.5}$ erg s$^{-1} > L_{5100} > 10^{46.6}$ erg s$^{-1}$), the contribution to $L_{5100}$ from host galaxy is negligible \citep{Shen11, Shen12}. This luminosity range complements the local $L_{5100}$-$L_{3000}$ relation at the brighter end.
 
In the literature, the continuum luminosity at 1350 \AA\ ($L_{1350}$) or the continuum luminosity at 3000 \AA\ ($L_{3000}$) is often used in the absence of $L_{5100}$ to estimate the BH masses based on UV spectra. The luminosities of broad emission lines are sometimes used, if the continuum is either too faint or contaminated by radiation from host galaxy or relativistic jets \citep{Wu04, Greene05, Shen11, Shen12}.  

We find strong correlations between $L_{5100}$ and broad line luminosities, i.e., $L_{\rm H\beta}$ and $L_{\rm MgII}$. Their Spearman correlation coefficients are 0.79$\pm$0.04 and 0.48$\pm$0.04, respectively, with confidence levels higher than 95\%. The correlation between $L_{3000}$ and $L_{5100}$ is strongest, yielding $r1$ of 0.84$\pm$0.01. These relations indicate that either luminosity can be used to estimate BH masses.

\section{A MODIFIED VIRIAL BH MASS ESTIMATE \label{Calibration}}
As shown in \textsection{4.1} and \textsection{4.2}, scatters in the line width correlations and the luminosity correlations can introduce uncertainties in BH mass estimates. Simultaneous observations of the MgII and H$\beta$ lines in our near-IR spectroscopy provide a unique opportunity to assess the reliability of the BH mass estimates for high redshift luminous quasars based on the MgII line.

We take the BH masses calculated based on FWHM$_{\rm H\beta}$ and $L_{5100}$ using the calibration from \cite{Vestergaard06} as the reference values. The virial BH mass estimates based on line width and luminosity can be expressed as:
\begin{equation} \label{eq:mbh}
\log(\frac{M_{\rm BH, vir}}{M_{\odot}}) = a + b \log (\frac{L}{10^{44} {\rm erg\ s^{-1}}}) + 
c \log {\rm (\frac{FWHM}{km\ s^{-1}})},
\end{equation}
where $L$ and FWHM are the continuum (line) luminosity and width of one specific line, respectively. Coefficients $a$, $b$ and $c$ are obtained from the linear regression analysis.

Modified calibrations for MgII are derived from the 19 quasars labeled as `good' with full coverage of both emission lines (H$\beta$ and MgII). We still assume that the broad line region gas is virialized. Thus, using the LINMIX$\_$ERR method \citep{Kelly07}, we fit the coefficients in Equation~\ref{eq:mbh} under Scheme 1, where $a$ and $b$ are free parameters and $c$ is fixed at 2.0. The LINMIX$\_$ERR method is a Bayesian approach to linear regression with error in one independent variable. For comparison, another fitting scheme (Scheme 2) is applied to Equation 1 using the MLINMIX$\_$ERR method \citep{Kelly07}; in this case, $a$, $b$ and $c$ are all free parameters. The MLINMIX$\_$ERR is a Bayesian approach to linear regression with errors in multiple independent variables. The derived coefficients and their uncertainties are the mean values and the standard deviations of their distributions from the Bayesian analysis, respectively. 

Based on the $z\sim3.5$ quasars in our sample, the result of Scheme 1 is as follows:
\begin{equation} \label{eq:mbh_mgii}
\log(\frac{M_{\rm BH, vir}}{M_{\odot}}) = (1.07\pm2.62) + (0.48\pm0.88) \log (\frac{L_{3000}}{10^{44}\ {\rm erg\ s^{-1}}}) + 
2 \log {\rm (\frac{FWHM_{MgII}}{km\ s^{-1}})}.
\end{equation}
While for Scheme 2, the coefficients $a$, $b$ and $c$ are 4.95$\pm$1.67, 0.25$\pm$0.31 and 1.11$\pm$0.30, respectively. 
These two modified calibrations are obtained from simultaneous observations of MgII and H$\beta$ lines for 19 high redshift luminous quasars. However, given the small size and the narrow luminosity range of our sample, the uncertainties in the best-fit parameters are relatively large. Thus we note that the modified calibration may not be the best for other studies. Here, the MgII based BH masses for our sample are estimated using Eq.~(\ref{eq:mbh_mgii}) and listed in Col. (7) of Table~\ref{fitting-results}.

\section{DISCUSSION\label{Discussion}}
\subsection{Comparison with Earlier Studies \label{comparison}}
For the 19 quasars labeled as `good' and with full coverage of both the two emission lines (H$\beta$ and MgII), based on the reliable reference BH mass estimates using the H$\beta$ line, we find that the BH masses are between $(1.90 \pm 0.24) \times 10^{9} M_{\odot}$ and $(1.37 \pm 0.04) \times 10^{10} M_{\odot}$. The median value is $5.14 \times 10^{9} M_{\odot}$ with the 1$\sigma$ dispersion $\sim$0.25 dex.

The reference BH masses are compared with $M_{\rm BH}$ estimated from the MgII emission line based calibrations given in earlier studies, and the distributions of the offsets are shown as the black histograms in Fig.~\ref {MBH_total1}. The red histogram refers to the distribution of the offsets between $\log M_{\rm BH}$ measured for our sample based on Eq.~(\ref{eq:mbh_mgii}) and the reference $\log M_{\rm BH}$. 
The MgII line-based BH masses estimated with Eq. (2) range from $(1.01 \pm 0.69) \times 10^{9} M_{\odot}$ to $(1.53 \pm 0.32) \times 10^{10} M_{\odot}$ with a median value of $3.37 \times 10^9 M_{\odot}$. The 1 $\sigma$ scatter of these $M_{\rm BH}$ estimates away from the median value is $\sim$ 0.37 dex. The $\log M_{\rm BH}$ uncertainties shown in Col. (7) of Table 2 only account for measurement uncertainties of FWHM$_{\rm MgII}$ and $L_{3000}$, through a Monte-Carlo approach mentioned in Section 3. The measurement uncertainties of the MgII line-based $\log\ M_{\rm BH}$ are between 0.02 dex and 0.46 dex, with a median value of 0.07 dex. These measurement uncertainties can be significantly amplified, after including the large uncertainties in the fitted parameters in Eq. (2).

Despite of large uncertainties in the best-fit parameters, the two newly derived calibrations from Scheme 1 and Scheme 2 still provide consistent BH mass estimates compared with the reference masses. The mean offsets from the reference $\log M_{\rm BH}$ are $-0.01\pm0.27$ and $-0.01\pm0.18$, respectively. The earlier studies and their calibrations based on FWHM$_{\rm MgII}$ and $L_{3000}$ are listed in the first four columns of Table~\ref{mbhdetail0}. The last two columns show the comparison results, i.e., the mean offset of the estimated $\log M_{\rm BH}$ values from the reference $\log M_{\rm BH}$ and the 1$\sigma$ dispersion of the deviation from the mean offset, respectively. In general, BH masses estimated with the calibrations from these earlier studies are consistent with the reference H$\beta$ based virial BH masses. 
Mean absolute offsets shown in Col. (6) of Table~\ref{mbhdetail0} are between $(0.01\pm0.23)$ dex and $(0.41\pm0.28)$ dex, with a mean value of 0.19 dex and 1$\sigma$ dispersion as 0.26 dex. The largest mean offset as $(0.41\pm0.28)$ dex comes from the mean offset between the $M_{\rm BH}$ estimated with the MgII line-based calibration \citep{Kollmeier06} and the reference $M_{\rm BH}$. We suspect that the larger offsets may be due to the sample selection; the sample that \cite{Kollmeier06} adopted covers relatively lower and narrower redshift (luminosity) ranges. The mean values of the offset distributions are also shown as the vertical dashed lines in Fig.~\ref{MBH_total1}.

If only the offsets are considered, our MgII based calibration show smaller offsets from the reference $\log M_{\rm BH}$, with the mean absolute offset value around $\sim$ 0.01 dex compared to the mean offset $\sim$ 0.19 dex for the earlier studies. However, after taking into account the relatively large 1$\sigma$ dispersions of these mean differences, our results generally agree with the calibrations of earlier studies. 

Thanks to the available SDSS spectroscopy of the $z \sim 3.5$ quasars, we also estimate BH masses using the FWHM$_{\rm CIV}$ and $L_{1350}$ based calibration from \cite{Vestergaard06}. The measured BH masses are compared with the reference BH masses and the MgII line-based $M_{\rm BH}$ estimates using Eq. (~\ref{eq:mbh_mgii}). They are generally consistent with each other, with mean offsets of $0.08\pm0.36$ dex and $0.10\pm0.33$ dex, respectively. 

In all, with simultaneous observations of the MgII and H$\beta$ lines, we ensure the reliability to measure BH masses based on the MgII line, without the influences of flux variability. Moreover, we get the consistent results with the earlier studies which implies that flux variability has marginal influence on BH mass measurements using the MgII line compared to the H$\beta$-based BH mass estimates.

\subsection{Black Hole Accretion and Growth}
Reliable estimates of black hole masses and Eddington ratios ($R_{\rm EDD} = {L_{\rm bol}} / {L_{\rm EDD}}$) of our sample can be obtained based on their H$\beta$ lines. Here the bolometric luminosity $L_{\rm bol}$ is calculated as $f_{\rm L} L_{5100}$ and the bolometric correction factor $f_{\rm L}$ is assumed to be 9.26 \citep{Richards06}. This provides a good laboratory to investigate the BH growth for quasars at $z \sim 3.5$. 

For the 20 quasars labeled as `good' with full H$\beta$ coverage, the minimum and maximum $R_{\rm EDD}$ values are $\sim0.30\pm0.01$ and $3.05\pm0.25$, respectively. The median value is 1.12 with 1 $\sigma$ scatter of 0.79. This indicates that these quasars are accreting at high $R_{\rm EDD}$ at this redshift (and luminosity) range. From the $\log R_{\rm EDD}$ measurements and their uncertainties shown in Col. (9) of Table 2, we notice that $R_{\rm EDD}$ of 5 quasars are above unity with 3$\sigma$ significance. If the typical systematic uncertainty of RM-based $M_{\rm BH}$ estimates as 0.3-0.4 dex \citep{Vestergaard02, Onken04, Peterson04} is considered, the $R_{\rm EDD}$ values of all the quasars are consistent with unity within 3 times their measurement uncertainties.

To investigate the BH growth time for the high redshift luminous quasars, we assume that the quasars are accreting at a constant $R_{\rm EDD}$ as listed in Table~\ref{fitting-results}. Then the growth time can be described similar to \cite{Netzer07}:
\begin{equation}
t_{\rm growth} = 0.38 {\rm Gyr} \ \frac{\eta/(1-\eta)}{f_{\rm L} L_{5100}/L_{\rm EDD}} \log {\frac{M_{\rm BH}}{M_{\rm seed}}}\frac{1}{f_{\rm active}},
\end{equation}
where the accretion efficiency $\eta$ is assumed to be 0.1 and $f_{\rm active}$ is the duty cycle of quasar activity, i.e., fraction of time when a black hole is active.

According to the current paradigm, the BH growth is due to accretion of surrounding gas onto much smaller initial seeds. BH seeds with tens to hundreds of $M_{\odot}$ and those with $10^4$ to $10^6 M_{\odot}$ can be traced back to the first generation stars \citep{Bromm99} and direct collapse of supermassive objects \citep{Begelman06}, respectively.   

With $f_{\rm active}$ of 1 and $M_{\rm seed}$ of $10^{4} M_{\odot}$, all the quasars in our sample can grow to their estimated BH masses within the age of the Universe at their corresponding redshifts. Compared to the $2.2<z<3.4$ quasars with $10^{45.2}{\ \rm erg\ s^{-1}}<L_{5100}<10^{46.4}$ erg s$^{-1}$ in \cite{Netzer07}, under the same settings of $f_{\rm active}$ and $M_{\rm seed}$, only 27\% of their 15 sources have enough time to grow their BH masses. For the quasars having not enough time to grow their BH masses at $z<3$, \cite{Netzer07} suggested that they may have gone through one or more past episodes with high accretion rates than the estimated values; the BH growth of the $z\sim3.5$ quasars in our sample seem to fit the scenario. Of course, our quasars are not likely from the same progenitor population of the quasars at $z < 3$ studied in \cite{Netzer07}, because at the observed accretion rates, the quasars in our sample already have larger BH masses than the quasars in their sample. 

As our sample is more luminous ($10^{46.6}{\ \rm erg\ s^{-1}}<L_{5100}<10^{47.5}$ erg s$^{-1}$) with higher accretion rates, given the simple BH growth model, it is not surprising to find that a larger fraction of quasars fits the growth scenarios at the same $f_{\rm active}$ and $M_{\rm seed}$. To investigate whether or not the systematic uncertainty of $M_{\rm BH}$ estimates influence the conclusion, we adopt a Monte-Carlo method to resample the BH mass of each quasar in our sample according to the assumed BH mass distribution; for each quasar the distribution of $\log M_{\rm BH}$ is a Gaussian with a peak at its $\log M_{\rm BH}$ value shown in Col. (6) of Table~\ref{fitting-results} and a dispersion of 0.3 dex (the typical systematic uncertainty of RM-based $M_{\rm BH}$ estimates). This process is iterated 50 times to generate 50 different mock samples with the same number of quasars as our sample. We find that the conclusion is not influenced by the systematic uncertainties of $M_{\rm BH}$ estimates; a larger fraction of quasars in our sample fit the growth scenario at the same $f_{\rm active}$ and $M_{\rm seed}$ than the sample in \cite{Netzer07}.

\section{CONCLUSION\label{Conclusion}}
In this paper we have empirically determined the relations between single epoch virial BH mass 
estimates based on different lines for 21 high redshift luminous quasars ($3.2 < z < 3.9$, $L_{5100} > 10^{46.6}$ erg s$^{-1}$), using high quality near-IR spectroscopy. Our sample has negligible
contamination from host galaxy, and is relatively large enough to obtain some statistically significant  
results for the high redshift luminous quasars. The main conclusions are
 as follows:\\
1. The MgII FWHM is well correlated with the H$\beta$ FWHM, confirming itself as a good substitute for FWHM$_{\rm H\beta}$ in the black hole mass estimates. \\
2. The continuum luminosity at 5100 \AA\ correlates well with the continuum luminosity at 3000 \AA\ and broad emission line luminosities (H$\beta$ and MgII).  \\
3. With simultaneous near-IR spectroscopy of the H$\beta$ and MgII lines, we ensure the reliability to estimate black hole masses based on the MgII line for high redshift quasars, without the influences of flux variability.  \\
4. With the reliable H$\beta$ line based black hole mass and Eddington ratio estimates, we find $1.90 \times10^{9} M_{\odot} \lesssim M_{\rm BH} \lesssim 1.37\times10^{10} M_{\odot}$ with a median of $\sim 5.14\times10^{9} M_{\odot}$. We also find that the $z\sim3.5$ quasars in our sample are accreting at Eddington ratios in the range from 0.30 to 3.05, with a median value of $\sim1.12$.  \\
5. With a duty cycle of 1 and a seed black hole mass at $10^{4} M_{\odot}$, the quasars in this $z\sim3.5$ sample can grow to their estimated black hole masses within the age of the Universe at their redshifts, under their high accretion rates. \\

We thank the anonymous referee for suggestions that have significantly improved the quality of this paper. We thank Yue Shen for helpful discussions and providing their IDL packages for our use, and Zhao-Yu Li for reading the manuscript and providing various helpful comments.
This work is supported by the China Scholarship Council (Wenwen Zuo), NSFC grants 
11373008 and 11033001. X. Fan, F. Bian, and R. Green acknowledge support from NSF grants AST 08-06861 and AST 11-07682. 
This research uses data obtained through the Telescope Access Program (TAP), which 
is funded by the National Astronomical Observatories, Chinese Academy of Sciences, 
and the Special Fund for Astronomy from the Ministry of Finance.
Observations obtained with the Hale Telescope at Palomar Observatory were obtained 
as part of an agreement between the National Astronomical Observatories, Chinese Academy
of Sciences, and the California Institute of Technology.
The LBT is an international collaboration among institutions in the United States, 
Italy and Germany. LBT Corporation partners are: The University of Arizona on behalf 
of the Arizona university system; Istituto Nazionale di Astrofisica, Italy; LBT 
Beteiligungsgesellschaft, Germany, representing the Max-Planck Society, the 
Astrophysical Institute Potsdam, and Heidelberg University; The Ohio State University, 
and The Research Corporation, on behalf of The University of Notre Dame, University 
of Minnesota and University of Virginia.
Funding for the SDSS and SDSS-II has been provided by the Alfred P.
Sloan Foundation, the Participating Institutions, the National Science
Foundation, the U.S. Department of Energy, the National Aeronau- tics
and Space Administration, the Japanese Monbukagakusho, the Max Planck
Society, and the Higher Education Funding Council for England. The
SDSS Web Site is http://www.sdss.org/. 
The SDSS is managed by the Astrophysical Research Consortium for the
Participating Institutions. The Participating Institutions are the
American Museum of Natural History, Astrophysical Institute Potsdam,
University of Basel, University of Cambridge, Case Western Reserve
University, University of Chicago, Drexel University, Fermilab, the
Institute for Advanced Study, the Japan Participation Group, Johns
Hopkins University, the Joint Institute for Nuclear Astrophysics, the
Kavli Institute for Particle Astrophysics and Cosmology, the Korean
Scientist Group, the Chinese Academy of Sciences (LAMOST), Los Alamos
National Laboratory, the Max-Planck-Institute for Astronomy (MPIA),
the Max- Planck-Institute for Astrophysics (MPA), New Mexico State
University, Ohio State University, University of Pittsburgh,
University of Portsmouth, Princeton University, the United States
Naval Observatory, and the University of Washington.

\bibliography{reference3_v1}

\begin{deluxetable}{cccccccccccccccccccc}
\tabletypesize{\tiny}
\tablecaption{The $z\sim3.5$ Sample \label{sample}}
\tablewidth{0pt}
\tablehead{
\colhead{Name (SDSS)}    & \colhead{$z$} &
\colhead{$i_{\rm PSF}$ (mag)} & \colhead{$J_{\rm 2MASS}$ (mag)}   & \colhead{$H_{\rm 2MASS}$ (mag)}  
& \colhead{$K_{\rm 2MASS}$ (mag)}  & \colhead{Exposure (s)} & \colhead{Obs. UT} \\
\colhead{(1)}      & \colhead{(2)}  & \colhead{(3)} & \colhead{(4)}  &
\colhead{(5)}      & \colhead{(6)}  & \colhead{(7)} & \colhead{(8)}
}
\startdata
 J011521.20+152453.3$^{\rm g}$ & 3.443&  18.424&  16.980&  16.478&  15.793   & 3600  & 111022  \\
 J012403.77+004432.6$^{\rm p}$ & 3.834&  17.907&  16.799&  15.908&  15.710   & 2400  & 111022  \\
 J014049.18-083942.5$^{\rm gb}$& 3.717&  17.521&  15.952&  15.572&  15.086   & 3600  & 111020  \\
 J014214.75+002324.2$^{\rm g}$ & 3.379&  17.948&  16.754&  16.242&  15.648   & 3600  & 111021  \\
 J015048.83+004126.2$^{\rm gb}$& 3.701&  18.433&  16.788&  16.305&  15.619   & 3600  & 111020  \\
 J015741.57-010629.6$^{\rm g}$ & 3.572&  18.062&  16.612&  16.151&  15.434   & 3600  & 111021  \\
 J021646.94-092107.2$^{\rm mb}$& 3.732&  17.871&  16.983&  16.105&  15.856   & 3600  & 111020/111022  \\
 J025021.76-075749.9$^{\rm g}$ & 3.337&  17.993&  16.661&  16.111&  15.853   & 3600  & 111021  \\
 J025905.63+001121.9$^{\rm g}$ & 3.373&  17.749&  16.192&  15.462&  15.187   & 3600  & 111021  \\
 J030341.04-002321.9$^{\rm g}$ & 3.233&  17.531&  16.104&  15.771&  15.225   & 3600  & 111022  \\
 J030449.85-000813.4$^{\rm g}$ & 3.287&  17.465&  16.304&  15.682&  15.286   & 3600  & 111020  \\
 J075303.34+423130.8$^{\rm g}$ & 3.590&  17.769&  16.698&  16.151&  15.162   & 3600  & 111020  \\
 J075819.70+202300.9$^{\rm m}$ & 3.761&  18.225&  16.951&  16.007&  15.702   & 3600  & 111021  \\
 J080430.56+542041.1$^{\rm g}$ & 3.759&  17.964&  16.599&  16.457&  15.348   & 3600  & 111022  \\
 J080819.69+373047.3$^{\rm g}$ & 3.480&  18.395&  16.780&  16.038&  15.624   & 3600  & 111021  \\
 J080956.02+502000.9$^{\rm g}$ & 3.281&  17.955&  16.978&  15.772&  15.345   & 3600  & 111022  \\
 J081855.77+095848.0$^{\rm g}$ & 3.700&  17.873&  16.710&  16.211&  15.788   & 3600  & 111021/111022  \\
 J084401.95+050357.9$^{\rm gb}$& 3.360&  17.089&  15.386&  14.928&  14.190   & 3600  & 111022  \\
 J090033.50+421547.0$^{\rm g}$ & 3.290&  16.678&  15.355&  14.668&  14.054   & 2400  & 120415  \\
 J094202.04+042244.5$^{\rm g}$ & 3.276&  17.176&  15.888&  15.325&  14.622   & 3600  & 120415  \\
 J102325.31+514251.0$^{\rm m}$ & 3.477&  17.599&  16.296&  15.841&  15.367   & 3600  & 120416  \\
 J115954.33+201921.1$^{\rm g}$ & 3.426&  17.076&  15.776&  15.369&  15.139   & 2400  & 120416  \\
 J121027.62+174108.9$^{\rm mb}$& 3.477&  17.719&  16.001&  15.598&  14.889   & 3600  & 120416  \\
 J150332.17+364118.0$^{\rm gb}$& 3.261&  17.361&  15.872&  15.667&  15.010   & 2400  & 120415  \\
 J173352.23+540030.4$^{\rm g}$ & 3.432&  17.120&  15.869&  15.724&  14.953   & 2400  & 120415  \\
 J213023.61+122252.0$^{\rm g}$ & 3.272&  17.929&  16.805&  15.889&  15.285   & 3600  & 111021  \\   
 J224956.08+000218.0$^{\rm g}$ & 3.311&  18.366&  16.774&  15.523&  14.816   & 3600  & 111022  \\
 J230301.45-093930.7$^{\rm g}$ & 3.492&  17.597&  16.773&  16.156&  15.640   & 3600  & 111020  \\
 J232735.67-091625.6$^{\rm p}$ & 3.263&  18.351&  16.962&  16.428&  16.061   & 3600  & 111021  \\
 J234625.66-001600.4$^{\rm m}$ & 3.507&  17.701&  16.873&  16.133&  15.302   & 3600  & 111020  \\
 J074521.78+734336.1$^{\rm m}$ & 3.220&  16.310&  15.064&  14.593&  13.943   & 3600  & 121211  \\ 
 J082535.19+512706.3$^{\rm g}$ & 3.512&  17.938&  16.212&  15.722&  15.002   & 2400  & 121211  \\
\enddata
\tablecomments{
Col. (1) Name of the quasars. `g', `m' and `p' refer to spectra labeled with `good', `median' and `poor', 
respectively. `b' means that the quasars are BALs. The last two targets were observed with the LBT.
Col. (2) Redshift measured from their [OIII] $\lambda 5007$ line of the near-IR spectra.
Col. (3) SDSS $i$-band PSF magnitudes.
Col. (4-6) 2MASS magnitudes.
Col. (7) Total integration time in each band for each target.
Col. (8) UT dates of near-IR observations.
}
\end{deluxetable} 

\begin{deluxetable}{cccccccccccccccccccc}
\tabletypesize{\tiny} 
\rotate
\tablecaption{The continuum and emission line parameters\label{fitting-results}}
\tablewidth{0pt}
\tablehead{
\colhead{Name (SDSS)}   &  \colhead{$\log L_{3000}$}  & \colhead{FWHM$_{\rm MgII}$} &
\colhead{$\log L_{5100}$}  & \colhead{FWHM$_{\rm H\beta}$} &
\colhead{$\log \rm M_{BH} (VP06)$} & \colhead{$\log M_{\rm BH}$}  &   
\colhead{$\log L_{\rm bol}$} & \colhead{$\log R_{\rm EDD}$}  &
\colhead{S/N} & \colhead{MgII}  & \colhead{H$\beta$} \\
\colhead{}       & \colhead{(erg s$^{-1}$)}   & \colhead{(km s$^{-1}$)} &
\colhead{(erg s$^{-1}$)}   & \colhead{(km s$^{-1}$)}  &
\colhead{($M_\odot$)}  & \colhead{($M_\odot$)} &  \colhead{(erg s$^{-1}$)}   &   
\colhead{}   & \colhead{}  & \colhead{} & \colhead{}  & \colhead{}\\
\colhead{(1)}   & \colhead{(2)}   & \colhead{(3)}   & \colhead{(4)}  & \colhead{(5)}  &
\colhead{(6)}   & \colhead{(7)}   & \colhead{(8)}   & \colhead{(9)}  & 
\colhead{(10)} & \colhead{(11)} & \colhead{(12)}}
\startdata
 J011521.20+152453.3$^{\rm g}$ &   46.70 $\pm$  $<0.01$  &  3363  $\pm$   326  &  46.61   $\pm$  0.01    &  5008  $\pm$  1237  &   9.61 $\pm$  0.26 &  9.43$\pm$  0.09     & 47.57$\pm$0.01     &  $-0.14\pm$ 0.25 &  14  & 1 & 1  \\ 
 J014214.75+002324.2$^{\rm g}$ &   46.86 $\pm$  $<0.01$  &  4700  $\pm$   421  &  46.61   $\pm$  $<0.01$ &  4796  $\pm$   619  &   9.58 $\pm$  0.11 & 9.80$\pm$  0.08   & 47.58$\pm$0.02    &  $-0.10\pm$ 0.08 &  22   & 1 & 1\\ 
 J015741.57-010629.6$^{\rm g}$ &   46.91 $\pm$  $<0.01$  &  5728  $\pm$   482  &  46.72   $\pm$  0.01   &  6692  $\pm$   534  &   9.92 $\pm$  0.06 & 9.99$\pm$  0.06   & 47.68$\pm$0.01    &  $-0.34\pm$ 0.07 &  50   & 1 & 1  \\ 
 J025021.76-075749.9$^{\rm g}$ &   46.85 $\pm$  $<0.01$  &  3332  $\pm$   310  &  46.63   $\pm$  $<0.01$ &  4071  $\pm$   282  &   9.44 $\pm$  0.06 & 9.49$\pm$  0.08   & 47.60$\pm$ $<0.01$ &  0.05 $\pm$ 0.05 &  10   & 1 & 1  \\ 
 J025905.63+001121.9$^{\rm g}$ &   47.10 $\pm$  $<0.01$  &  2913  $\pm$   325  &  47.01   $\pm$  $<0.01$ &  4482  $\pm$   310  &   9.72 $\pm$  0.06 & 9.50$\pm$  0.07   & 47.97$\pm$ $<0.01$ &  0.16 $\pm$ 0.07 &  23   & 1 & 1  \\ 
 J030341.04-002321.9$^{\rm g}$ &   47.00 $\pm$  $<0.01$  &  2986  $\pm$   219  &  46.82   $\pm$  0.01    &  3010  $\pm$   209  &   9.28 $\pm$  0.05 & 9.47$\pm$  0.06   & 47.79$\pm$ 0.01    &  0.41 $\pm$ 0.05 &  51   & 1 & 1  \\ 
 J030449.85-000813.4$^{\rm g}$ &   47.03 $\pm$  $<0.01$  &  1959  $\pm$   109  &  46.82   $\pm$  0.01    &  3366  $\pm$   335  &   9.38 $\pm$  0.09 & 9.12$\pm$  0.05   & 47.79$\pm$ 0.01    &  0.31 $\pm$ 0.10 &  37   & 1 & 1  \\ 
 J075303.34+423130.8$^{\rm g}$ &   46.84 $\pm$  $<0.01$  &  6643  $\pm$   147  &  46.74   $\pm$  $<0.01$ &  8477  $\pm$   113 &   10.14 $\pm$ 0.01  & 10.09$\pm$  0.02  & 47.71$\pm$ $<0.01$ &  $-0.53\pm$ 0.01 &  35   & 1 & 1 \\ 
 J075819.70+202300.9$^{\rm m}$ &   46.83 $\pm$  0.01  &  4483  $\pm$   178  &  46.67   $\pm$  0.04    &  6596  $\pm$  1806  &   9.89 $\pm$  0.22 & 9.74$\pm$  0.03   & 47.64$\pm$ 0.04    &  $-0.35\pm$ 0.18 &  15   & 1 & 1  \\ 
 J080430.56+542041.1$^{\rm g}$ &   47.08 $\pm$  $<0.01$  &  6140  $\pm$   543  &  46.91   $\pm$  $<0.01$ &  3902  $\pm$   940  &   9.55 $\pm$  0.16 &10.13$\pm$  0.07   & 47.87$\pm$ $<0.01$    &  0.23 $\pm$ 0.17 &  33   & 1 & 1  \\ 
 J080819.69+373047.3$^{\rm g}$ &   46.93 $\pm$  $<0.01$  &  7069  $\pm$   640  &  46.67   $\pm$  0.01    &  7967  $\pm$   228  &  10.05 $\pm$  0.03 &10.18$\pm$  0.09   & 47.64$\pm$ 0.01    & $-0.51\pm$  0.03 & 37   & 1 & 1  \\ 
 J080956.02+502000.9$^{\rm g}$ &   46.86 $\pm$  $<0.01$  &  3433  $\pm$  1655  &  46.74   $\pm$  $<0.01$ &  5803  $\pm$   204  &   9.81 $\pm$  0.03 & 9.52$\pm$  0.40   & 47.70$\pm$ $<0.01$ & -0.20 $\pm$ 0.03 &  45   & 1 & 1  \\ 
 J081855.77+095848.0$^{\rm g}$ &   46.90 $\pm$  $<0.01$  &  6364  $\pm$   171  &  46.74   $\pm$  0.01    &  5528  $\pm$   473  &   9.77 $\pm$  0.07 &10.08$\pm$  0.02   & 47.71$\pm$ 0.01    &  $-0.16\pm$ 0.07 &  16   & 1 & 1  \\ 
 J090033.50+421547.0$^{\rm g}$ &   47.40 $\pm$  $<0.01$  &  3017  $\pm$    65  &  47.25   $\pm$  0.02    &  3534  $\pm$   168  &   9.63 $\pm$  0.04 & 9.67$\pm$  0.02   & 48.22$\pm$ 0.02    &  0.48 $\pm$ 0.04 &  80   & 1 & 1  \\ 
 J094202.04+042244.5$^{\rm g}$ &   47.14 $\pm$  $<0.01$  &  2292  $\pm$   205  &  47.03   $\pm$  0.01    &  4396  $\pm$   354  &   9.71 $\pm$  0.06 & 9.31$\pm$  0.08   & 48.00$\pm$ 0.01    &  0.18 $\pm$ 0.07 &  61   & 1 & 1  \\ 
 J102325.31+514251.0$^{\rm m}$ &   46.84 $\pm$  0.02     & 10978  $\pm$  1188  &  46.80   $\pm$  0.02    &  4335  $\pm$   732  &   9.58 $\pm$  0.14 &10.52$\pm$  0.08   & 47.77$\pm$ 0.02    &  0.08 $\pm$ 0.14 &  25   & 1 & 1  \\ 
 J115954.33+201921.1$^{\rm g}$ &   47.20 $\pm$  $<0.01$  &  5847  $\pm$   234  &  46.99   $\pm$  $<0.01$ &  6599  $\pm$   337  &  10.05 $\pm$  0.04 &10.15$\pm$  0.03   & 47.96$\pm$ 0.01    &  $-0.19\pm$ 0.05 &  18   & 1 & 1  \\ 
 J173352.23+540030.4$^{\rm g}$ &   47.15 $\pm$  $<0.01$  &  2941  $\pm$   179  &  46.90   $\pm$  $<0.01$ &  3738  $\pm$    54  &   9.51 $\pm$  0.01 & 9.53$\pm$  0.05   & 47.87$\pm$ $<0.01$ &  0.26 $\pm$ 0.01 &  37   & 1 & 1  \\ 
 J213023.61+122252.0$^{\rm g}$ &   46.85 $\pm$  $<0.01$  &  1904  $\pm$   645  &  46.74   $\pm$  $<0.01$ &  4256  $\pm$    90  &   9.54 $\pm$  0.02 &  9.00$\pm$  0.30  & 47.71$\pm$ $<0.01$ &  0.07 $\pm$ 0.02 &  30   & 1 & 1  \\ 
 J224956.08+000218.0$^{\rm g}$ &   46.98 $\pm$  $<0.01$  &  2322  $\pm$   988  &  46.95   $\pm$  0.01    &  3288  $\pm$   932  &   9.42 $\pm$  0.20 & 9.24$\pm$  0.46   & 47.92$\pm$ 0.01    &  0.40 $\pm$ 0.19 &  10   & 1 & 1  \\ 
 J230301.45-093930.7$^{\rm g}$ &   46.88 $\pm$  $<0.01$  &  5612  $\pm$   291  &  46.68   $\pm$  $<0.01$ &  5887  $\pm$   142  &   9.79 $\pm$  0.02 & 9.96$\pm$  0.04   & 47.64$\pm$ $<0.01$ &  $-0.25\pm$ 0.02 &  46   & 1 & 1  \\ 
 J234625.66-001600.4$^{\rm m}$ &   46.81 $\pm$  $<0.01$  &  1198  $\pm$    61  &  46.75   $\pm$  $<0.01$ &  3391  $\pm$   118  &   9.35 $\pm$  0.03 & 8.59$\pm$  0.04   & 47.72$\pm$ $<0.01$ &  0.27 $\pm$ 0.03 &  34   & 1 & 1  \\ 
 J074521.78+734336.1$^{\rm g}$ &   47.47 $\pm$  $<0.01$  &  5894  $\pm$    77  &  47.33   $\pm$  0.01    &         $-$         &         $-$          &10.29$\pm$  0.01  & 48.29$\pm$ 0.01    &  $-0.09\pm$ 0.01 &  38   & 1 & 0  \\                        
 J082535.19+512706.3$^{\rm g}$ &    $-$                  &         $-$         &  46.93   $\pm$  0.01    &  6918  $\pm$   342  &  10.05 $\pm$  0.037 &     $-$             & 47.89$\pm$0.01    &  $-0.26\pm$ 0.04 &  19   & 0 & 1  \\ 
\enddata
\tablecomments{
Col.(2) Continuum luminosity at 3000 \AA.
Col.(3) FWHM of the broad MgII component. 
Col.(4) Continuum luminosity at 5100 \AA.
Col.(5) FWHM of the broad H$\beta$ component. 
Col.(6) H$\beta$ line-based BH masses and their uncertainties estimated based on the calibration from \cite{Vestergaard06}. The uncertainties quoted are only from statistical errors and not including systematic uncertainties of BH mass calibration ($\sim0.3-0.4$ dex, \citep{Vestergaard02}).
Col.(7) MgII line-based BH masses and their uncertainties estimated based on Eq.~\ref{eq:mbh_mgii}, without considering the uncertainties of the fitting parameters ($a$ and $b$).
Col.(8) Bolometric luminosity $L_{\rm bol} = 9.26 L_{5100}$ \citep{Richards06}. 
Col.(9) Eddington ratio and their uncertainties calculated from Col. (8) and Col. (6) (Col.(7) is used if no full coverage of H$\beta$ is available).
Col.(10) Median S/N per spectral element of 3 pixels for the near-IR spectrum. 
Col.(11) 1 means full coverage of MgII. 
Col.(12) 1 means full coverage of H$\beta$. 
}
\end{deluxetable}

\begin{deluxetable}{cccccccccccccccccccc}
\tabletypesize{\scriptsize}
\tablecaption{Correlations of Spectral Properties \label{fwhm}}
\tablewidth{0pt}
\tablehead{
\colhead{}     &  \colhead{$r1$}   & \colhead{$p$}  & \colhead{$r2$}           
	&  \colhead{slope$_{\rm OLS}$} & \colhead{scatter (dex)} & \colhead{slope$_{\rm BCES}$}   \\
 \colhead{1}       & \colhead{(2)}   & \colhead{(3)}   & \colhead{(4)}  & \colhead{(5)}  &
\colhead{(6)}   & \colhead{(7)}}   
\startdata
FWHM$_{\rm MgII}$ vs. FWHM$_{\rm H\beta}$            &  0.72$\pm$0.10 & 0.00$\pm$0.01  &  0.75$\pm$0.11 & 1.09$\pm$0.23 &0.29 & 1.44$\pm$0.33\\
FWHM$_{\rm CIV}$ vs. FWHM$_{\rm H\beta}$             &  0.29$\pm$0.07 & 0.21$\pm$0.15  &  0.24$\pm$0.07 & 0.27$\pm$0.26 &0.40 & 1.10$\pm$0.14\\ 
$L_{5100}$ vs. $L_{3000}$        &  0.84$\pm$0.01  &  0.00$\pm$$<0.01$ &  0.93$\pm$0.01 & 0.89$\pm$0.09 & 0.07 & 0.96$\pm$ 0.17 \\
$L_{5100}$ vs. $L_{1350}$        &  0.59$\pm$0.01  &  0.01$\pm$$<0.01$ &  0.56$\pm$0.01 & 0.69$\pm$0.24 & 0.57 & 1.20$\pm$ 0.38 \\
$L_{5100}$ vs. $L_{\rm H\beta}$  &  0.79$\pm$0.04  &  0.00$\pm$$<0.01$ &  0.84$\pm$0.05 & 1.27$\pm$0.20 & 0.37 & 1.43$\pm$ 2.87 \\
$L_{5100}$ vs. $L_{\rm MgII}$    &  0.48$\pm$0.04   &  0.04$\pm$0.02 &  0.67$\pm$0.09 & 1.05$\pm$0.28 & 0.73 & 0.96$\pm$ 2.32 \\
$L_{5100}$ vs. $L_{\rm CIV}$     &  0.73$\pm$0.03  &  0.00$\pm$$<0.01$ &  0.80$\pm$0.03 & 0.99$\pm$0.17 & 0.30 & 1.22$\pm$ 1.81 \\
\enddata
\tablecomments{
Col. (1) $L_{\rm H\beta}$, $L_{\rm MgII}$ and $L_{\rm CIV}$ refer to luminosities of broad components of the H$\beta$, MgII and CIV lines, respectively.
Col. (2) Spearman rank correlation coefficient.
Col. (3) Probability of $r1$ deviating from the null hypothesis.
Col. (4) Pearson's correlation coefficient.
Col. (5) Slope from the fit of their logarithmic values using the OLS method.
Col. (6) Scatter perpendicular to the best-fitting linear relation using the OLS method.
Col. (7) Slope from the fit of their logarithmic values using the BCES method.
}
\end{deluxetable}

\begin{deluxetable}{cccccccccccccccccccc}
\tabletypesize{\scriptsize}
\tablecaption{$M_{\rm BH}$ from previous MgII based calibrations vs. the reference $M_{\rm BH}$ from H$\beta$ based calibration in \cite{Vestergaard06}\label{mbhdetail0}}
\tablewidth{0pt}
\tablehead{
\colhead{Ref.} & \colhead{$a$}  &  \colhead{$b$}  & \colhead{$c$} &  
$z$  &  \colhead{Mean offset (dex)} & \colhead{$\sigma$ (dex)}  \\
\colhead{(1)}   & \colhead{(2)}   & \colhead{(3)}   & \colhead{(4)}  & \colhead{(5)}  &
\colhead{(6)}   & \colhead{(7)}    } 
\startdata
MD04     & 0.505 & 0.620 & 2.000 &  0.1-2.1       & -0.17  & 0.27  \\
VO09      & 0.860 & 0.500 & 2.000 &  0.08-2.5   &  -0.17  & 0.27  \\
W09       & 2.710 & 0.460 & 1.480 &  0.4-0.8       & -0.30   & 0.21  \\
S12        & 1.816 & 0.584 & 1.712 &  1.5-2.2       &  0.01   & 0.23  \\ 
K06        & 0.310 & 0.880 & 2.000 &  0.4-0.8       &  0.41   & 0.28  \\
T12         & 0.748 & 0.620 & 2.000 &  1.0-5.0      &  0.08   & 0.27  \\ 
\enddata
\tablecomments{
Col. (1) References of previous calibrations: MD04 \citep{McLure04}, VO09 \citep{Vestergaard09}, 
W09 \citep{Wang09}, S12 \citep{Shen12}, K06 \citep{Kollmeier06} and T12 \citep{Trakhtenbrot12}.
Col. (2-4) The values of $a$, $b$ and $c$ in Eq.~\ref{eq:mbh}. 
Col. (5) The redshift range of the sample utilized in references shown in Col.(1).
Col. (6) The mean offset of the $\log M_{\rm BH}$ values estimated with the MgII line-based calibrations away from the reference $\log M_{\rm BH}$ obtained using the H$\beta$ line-based calibration in \cite{Vestergaard06}.
Col. (7) The 1$\sigma$ uncertainty of the deviation from the mean offset.
}
\end{deluxetable}

\clearpage

\begin{figure}
\plotone{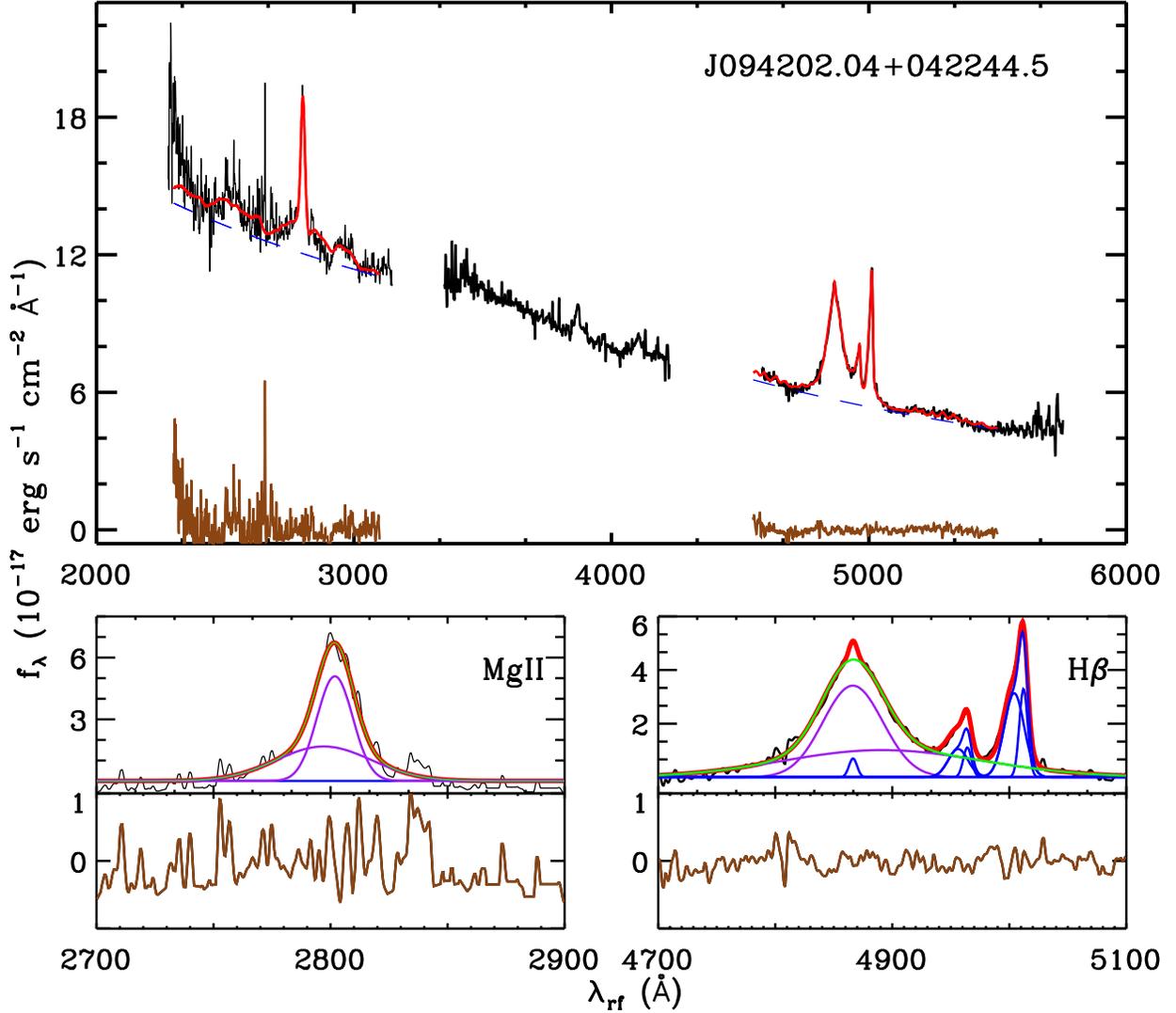}
\caption{An example of our model fits to the near-IR spectrum of SDSS J094202.04+042244.5. Here $\lambda_{\rm rf}$ refers to the rest-frame wavelength. 
Top panel: The observed spectrum is shown as the solid black line. The model (solid red line) 
is composed of power-law continuum (blue dashed line), FeII template and gaussian fits to emission lines, such as the MgII, H$\beta$ and [OIII] lines. 
Bottom panels: The red lines show the combined model fitting of MgII and H$\beta$. In each panel, the purple ones refer to the two gaussians of the broad component and their combined profile is shown in green; the blue lines represents the model fits for their narrow components and the brown lines show the fitting residuals.
\label{J0942}}
\end{figure}

\begin{figure}
\epsscale{0.85}
\plotone{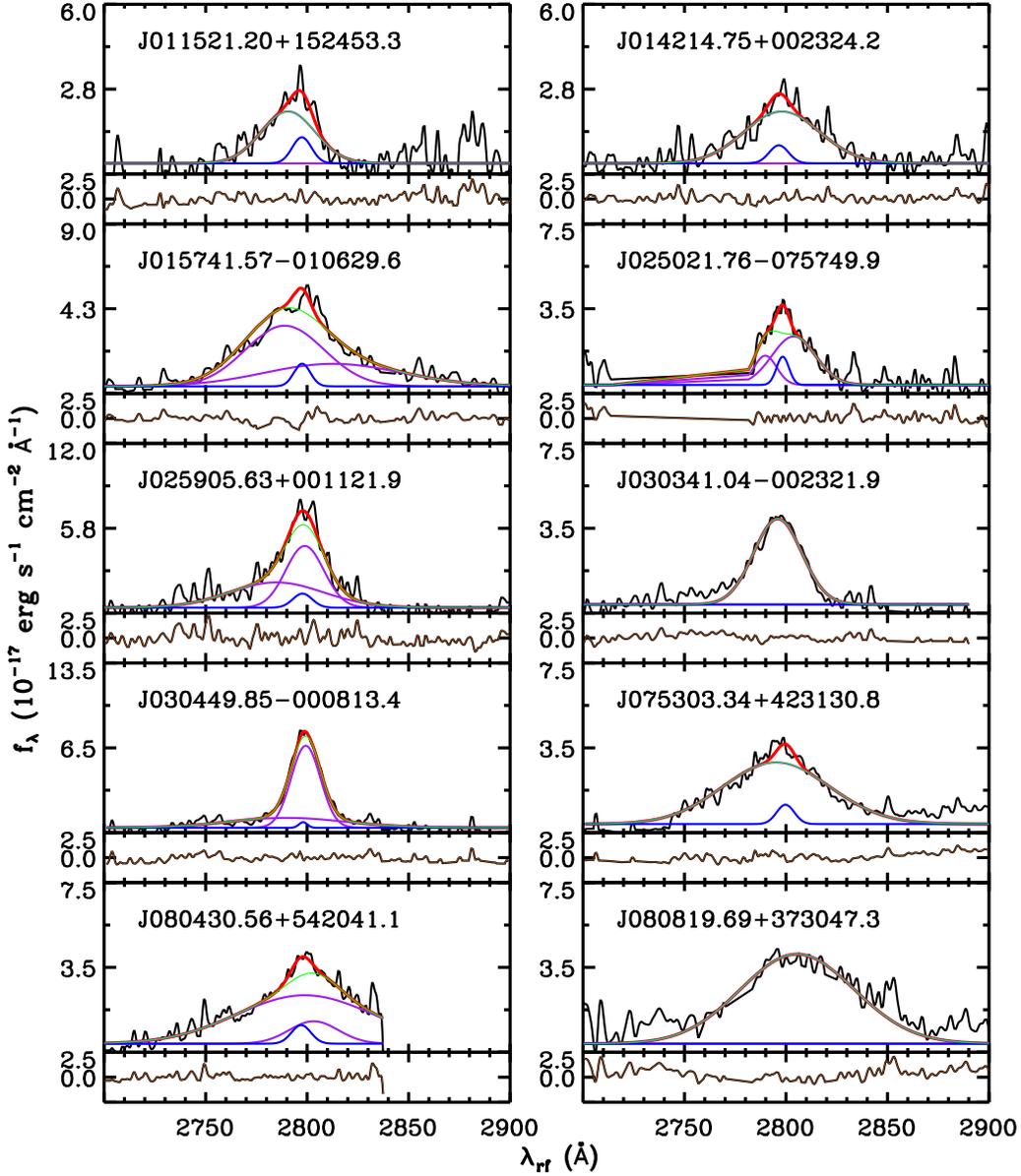}
\caption{Fitting results of the MgII lines for the 20 targets labeled as `good' and with full coverage of the MgII line. Note that the two targets (J0804 and J2130) have truncated profiles, but due to their acceptable MgII line fitting, we still take them as objects with full coverage of MgII. The spectrum in each panel is shown as the black line. The red lines show the combined model fitting of the emission line, where the purple ones refer to the two gaussians of the broad component and their combined profile is shown in green; the blue lines represent the model fits for the narrow components and the brown lines show the fitting residuals.
\label{allspec_0_MgII}}
\end{figure}

\setcounter{figure}{1}
\begin{figure}
\plotone{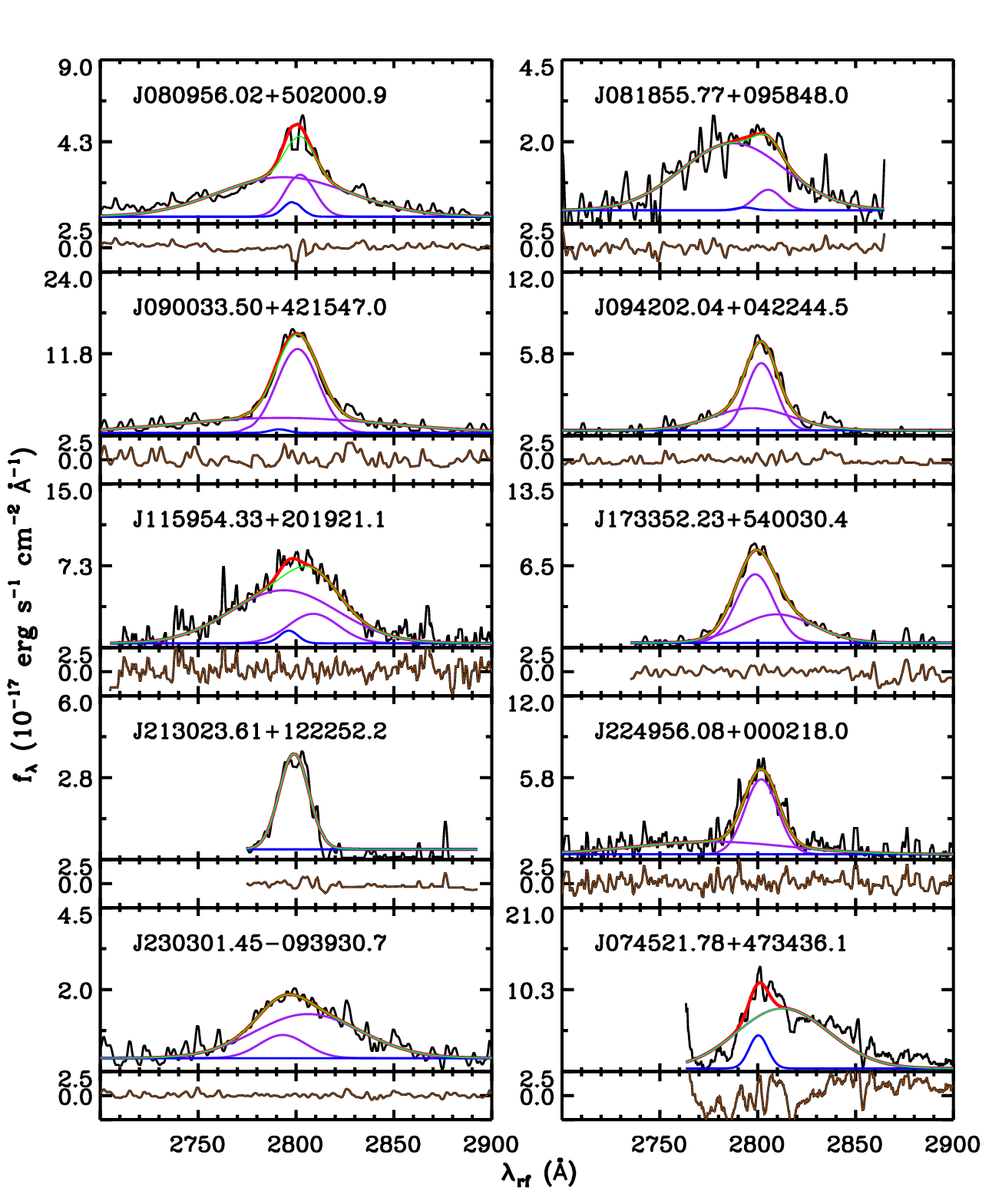}
\caption{Continued}
\end{figure}

\begin{figure}
\epsscale{0.88}
\plotone{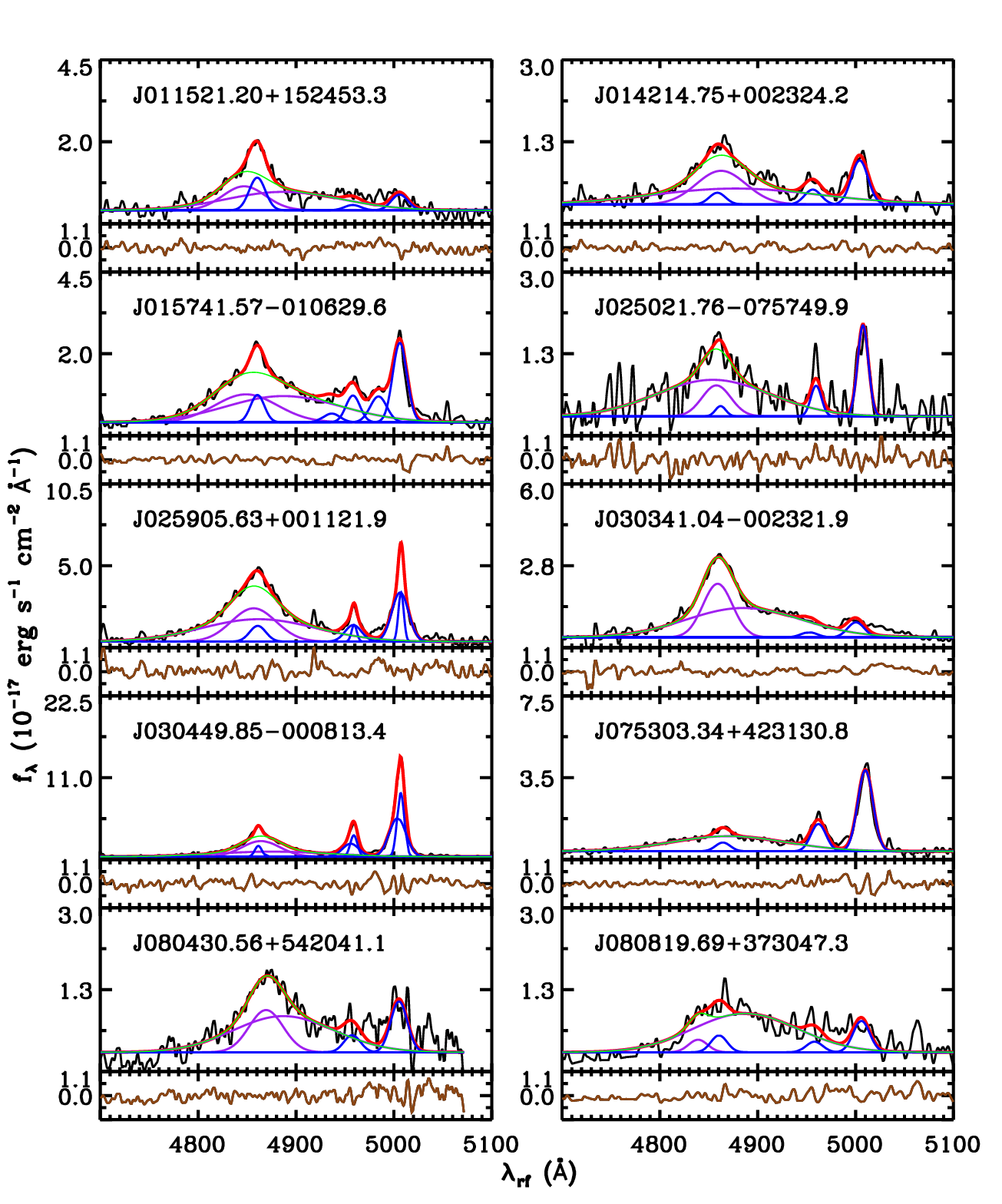}
\caption{Fitting results of the H$\beta$ lines for the 20 targets labeled as `good' and with full coverage of the H$\beta$ line. Colors of these lines have the same meaning with Fig.~\ref{allspec_0_MgII}.  
\label{allspec_0_HB}}
\end{figure}

\setcounter{figure}{2}
\begin{figure}
\plotone{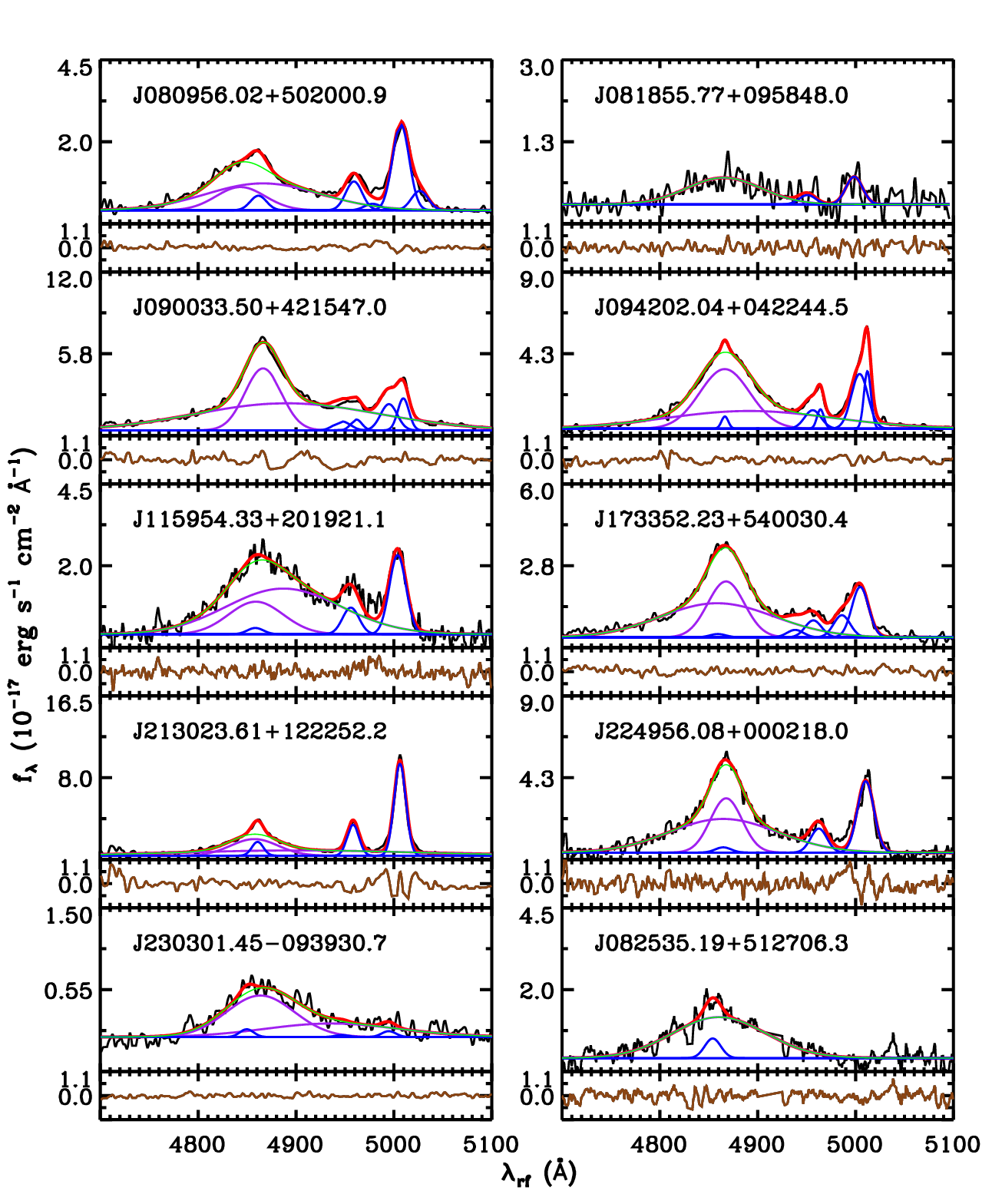}
\caption{Continued}
\end{figure}

\begin{figure}
\epsscale{0.88}
\plotone{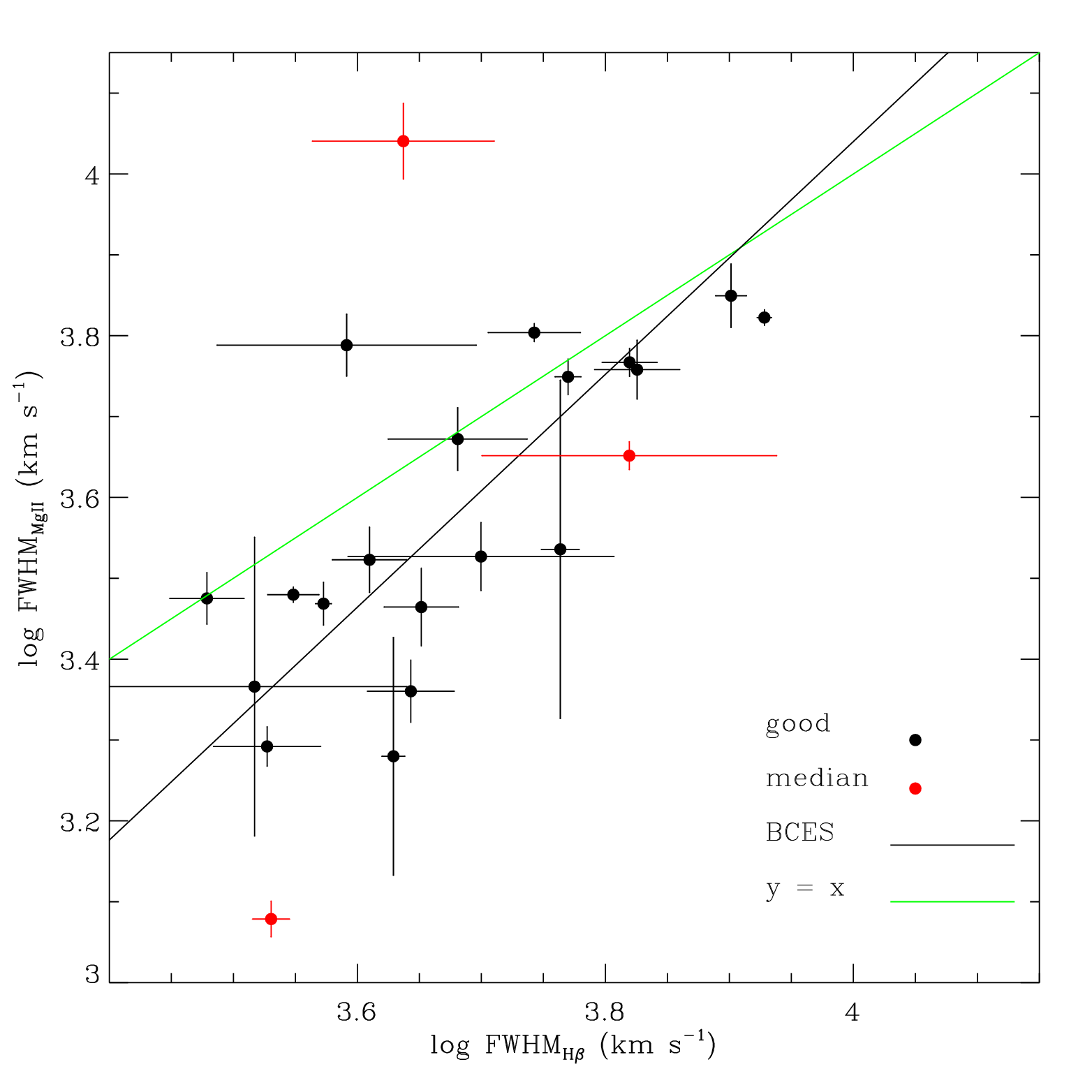}
\caption{$\log$ FWHM$_{\rm MgII}$ vs. $\log$ FWHM$_{\rm H{\beta}}$.
The red symbols refer to the quasars with spectra labeled as `median', the black ones 
represent the quasars labeled as `good'. The green line denotes the 1:1 relation, while the
black line refers to the fitting using the BCES method. 
\label{fwhm_total}}
\end{figure}

\begin{figure}
\plotone{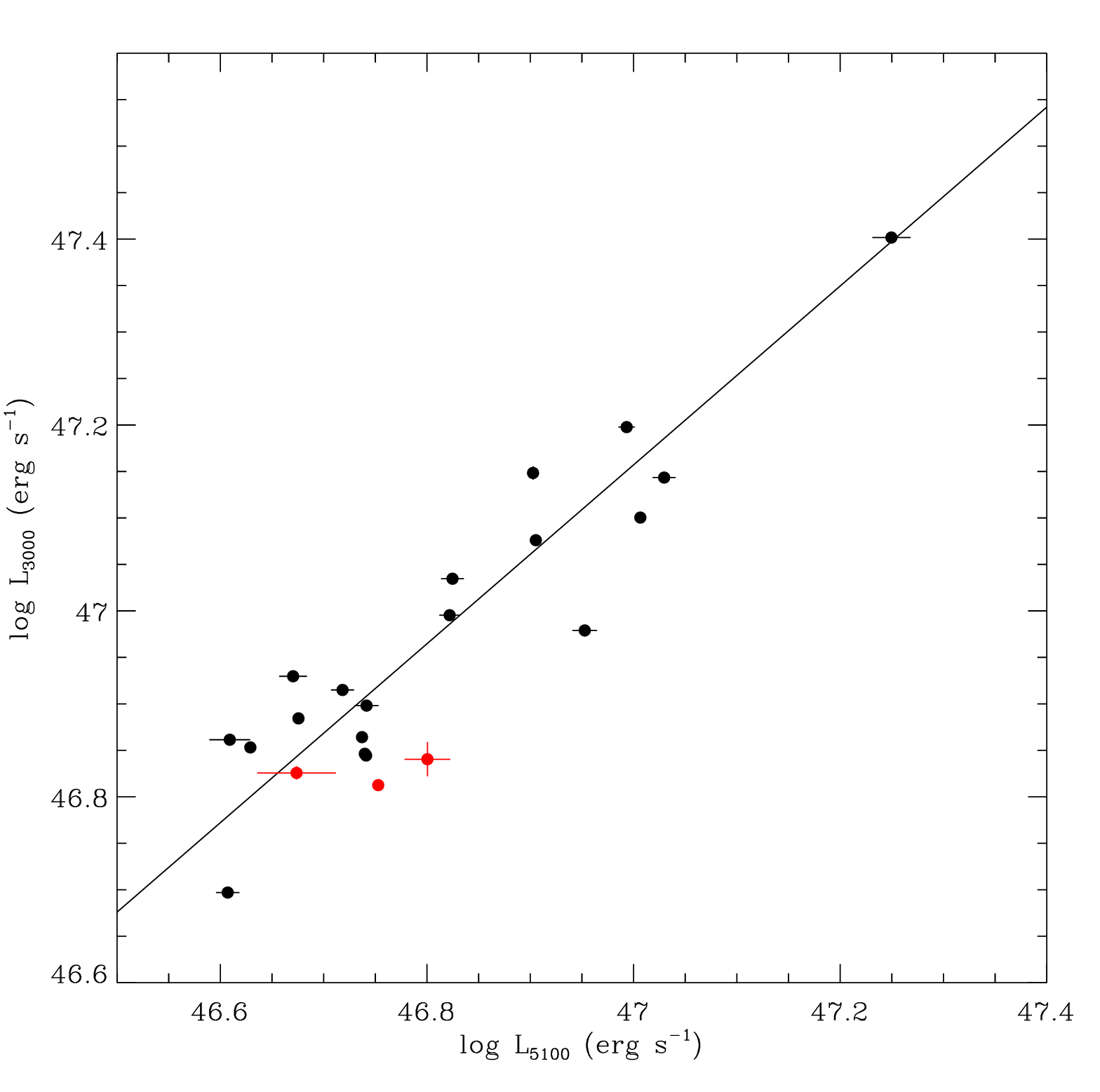}
\caption{$\log L_{3000}$ vs. $\log L_{5100}$. 
The red dots refer to the quasars with spectra labeled as `median', while the black ones respresent the quasars labeled as `good'. The black line denotes the fitting using the BCES method.
\label{l_total}}
\end{figure}

\begin{figure}
\plotone{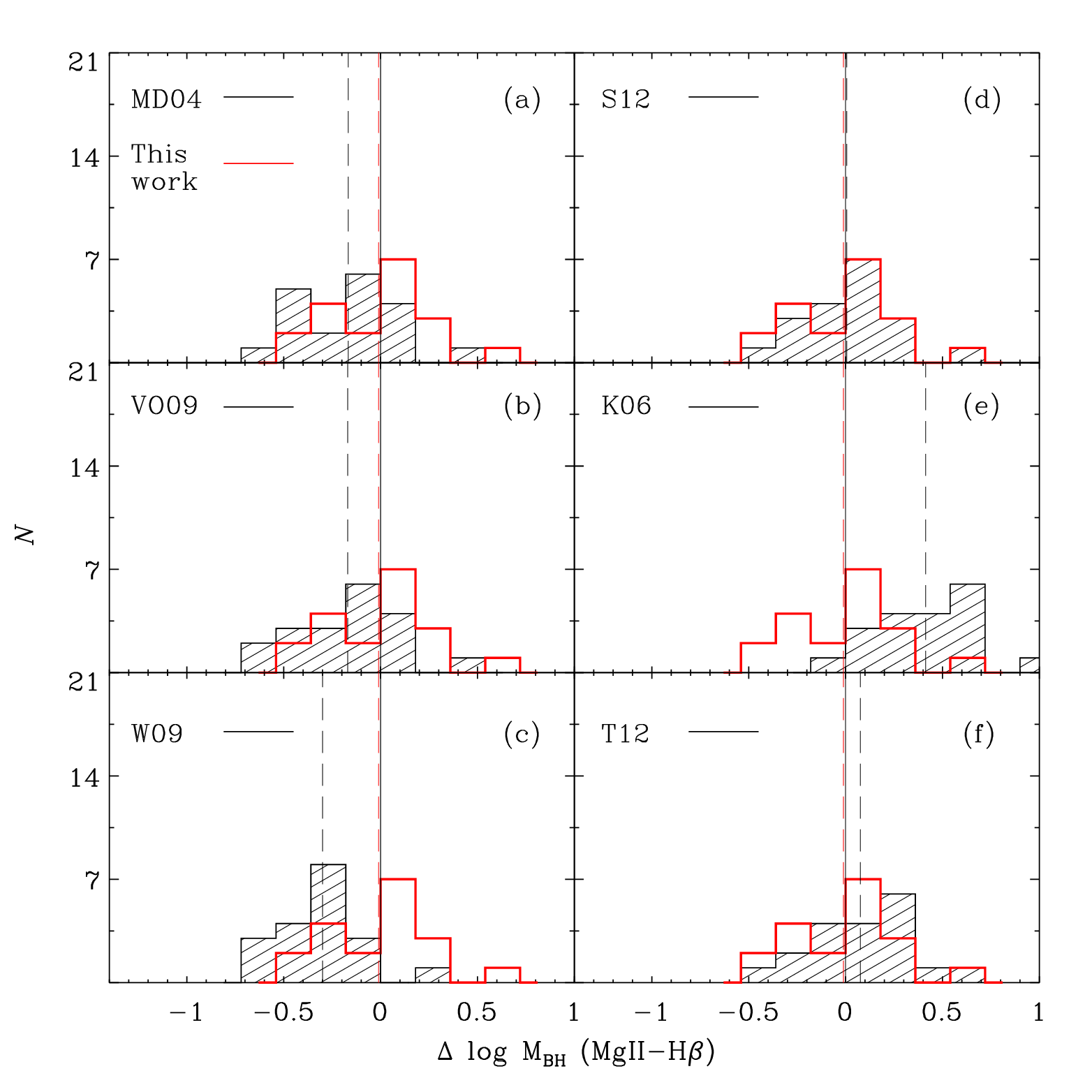}
\caption{The distributions of the offsets between different FWHM$_{\rm MgII}$ and $L_{3000}$ based $\log M_{\rm BH}$ and the reference $\log M_{\rm BH}$ calculated based on the FWHM$_{\rm H\beta}$ and $L_{5100}$ using the calibration from \cite{Vestergaard06}. The vertical dashed lines refer to the mean values of the offset distributions and the solid black line denotes the offset value at 0. The red histogram is the same in each panel.
\label{MBH_total1}} 
\end{figure}

\end{document}